\documentclass[a4paper,12pt]{article}
\usepackage{jheppub}
\usepackage[T1]{fontenc}
\usepackage{amsmath,amssymb,amsfonts,graphicx,subfigure,
slashed,amsthm,mathtools,braket,upgreek,bm,enumerate,tensor}
\usepackage[dvipsnames]{xcolor}
\usepackage{arydshln}
\usepackage{braket}
\usepackage{color}
\usepackage{hyperref}
\usepackage[utf8]{inputenc}
\usepackage[titletoc]{appendix}

\definecolor{phthaloblue}{rgb}{0.0, 0.06, 0.54}
\usepackage{hyperref}
\hypersetup{colorlinks=true, linkcolor=phthaloblue, citecolor=phthaloblue, urlcolor=phthaloblue}

\newcommand{\ie}{\textit{i.e}}
\usepackage{acro}
\DeclareAcronym{gr}{
    short=GR ,
    long=general relativity
}
\DeclareAcronym{uv}{
    short=UV ,
    long=ultraviolet
}
\DeclareAcronym{ir}{
    short=IR ,
    long=infrared
}
\DeclareAcronym{hl}{
    short=HL ,
    long=Ho\v{r}ava-Lifshitz
}
\DeclareAcronym{rg}{
    short=RG ,
    long=renormalization group
}
\DeclareAcronym{adm}{
    short=ADM ,
    long={Arnowitt, Deser and Misner}
}
\DeclareAcronym{flrw}{
    short=FLRW ,
    long={Friedmann-Lema\^{i}tre-Robertson-Walker}
}
\DeclareAcronym{pgws}{
    short=PGWs ,
    long={Primordial Gravitational Waves}
}

\title{\bf HBT Interferometry and Quantum Nature of 
Primordial Gravitational Waves in Ho\v{r}ava-Lifshitz Gravity 
}

\author{\large Sugumi Kanno ${}^{a,b}$, Hiroki Matsui ${}^b$, Shinji Mukohyama ${}^{b,c}$}

\emailAdd{kanno.sugumi@phys.kyushu-u.ac.jp}
\emailAdd{hiroki.matsui@yukawa.kyoto-u.ac.jp}
\emailAdd{shinji.mukohyama@yukawa.kyoto-u.ac.jp}
\affiliation{${}^a$Department of Physics, Kyushu University, 
Fukuoka 819-0395, Japan \medskip\\
${}^b$Center for Gravitational Physics and Quantum Information, Yukawa Institute for Theoretical Physics, Kyoto University, Kitashirakawa Oiwakecho, Sakyo-ku, Kyoto 606-8502, Japan \medskip\\
${}^{c}$Kavli Institute for the Physics and Mathematics of the Universe (WPI), The University of Tokyo Institutes for Advanced Study, The University of Tokyo, Kashiwa, Chiba 277-8583, Japan \medskip\\}

\abstract{
Ho\v{r}ava-Lifshitz gravity (to be precise, its projectable version) is recognized as a renormalizable, unitary, and asymptotically free quantum field theory of gravity. Notably, one of its cosmological predictions is that it can produce scale-invariant primordial density fluctuations and primordial gravitational waves without relying on inflation. In this paper, we investigate the quantum nature of the primordial gravitational waves generated in Ho\v{r}ava-Lifshitz gravity. It has been suggested that, for some inflationary models, the non-classicality of primordial gravitational waves in the squeezed coherent quantum state can be detected using the Hanbury Brown - Twiss (HBT) interferometry. We show that in Ho\v{r}ava-Lifshitz gravity, scale-invariant primordial gravitational waves can be generated during both the radiation-dominated and matter-dominated eras of the Universe. Moreover, the frequency range of their quantum signatures is shown to extend beyond that of inflationary models.}

\keywords{}
\preprint{YITP-24-160,  IPMU24-0044}

\begin{document}

\maketitle

\section{Introduction}

Attempts to construct a quantum field theory of gravity face many challenges. As is well known, \ac{gr} is recognized as being non-renormalizable, mainly because the Newtonian constant $G_N$ has dimensions (the mass dimension is $[G_N] = -2$). Non-renormalizability is known to cause uncontrollable \ac{uv} divergences. While adding higher curvature terms to the Einstein-Hilbert action can make the theory renormalizable~\cite{Stelle:1976gc}, it also risks generating massive ghosts, leading to a non-unitary quantum theory in the \ac{uv} regime.

To address these issues, \ac{hl} gravity theory~\cite{Horava:2009uw,Horava:2009if} has been proposed. This theory allows for renormalization based on power counting by including higher spatial curvature terms. The action and equations of motion of \ac{hl}  gravity include only up to second-order time derivatives, thus avoiding the issue of Ostrogradsky ghosts arising from higher-order time derivatives. Recently, the \ac{hl} theory (to be precise, its projectable version) has been demonstrated to be perturbatively renormalizable~\cite{Barvinsky:2015kil,Barvinsky:2017zlx} and asymptotically free~\cite{Barvinsky:2023uir,Barvinsky:2024svc}, and it is considered a viable \ac{uv} completion path for quantum gravity.

One of the fundamental features of \ac{hl} gravity is its \textit{anisotropic scaling}, also known as \textit{Lifshitz scaling}. In this scaling, the time coordinate $t$ and spatial coordinate vector $\vec{x}$ scale as $t \to b^z t$ and $\vec{x} \to b \vec{x}$ respectively, where $z$ is the \textit{dynamical critical exponent}. In $3+1$ dimensions, the anisotropic scaling with $z=3$ in the \ac{uv} breaks Lorentz symmetry but ensures renormalizability. The Lorentz symmetry is preserved in the \ac{ir} regime.
This theory has an outstanding cosmological prediction that scale-invariant primordial perturbations can be generated in any expansion of the early Universe $a \propto t^n$ with $n>1/3$~\cite{Mukohyama:2009gg}.
Additionally, this theory can potentially solve the horizon problem without inflation and offer a solution to the flatness problem through the so-called \textit{anisotropic instanton}~\cite{Bramberger:2017tid}. Furthermore, the \ac{hl} gravity addresses perturbation issues associated with the DeWitt boundary condition in quantum cosmology~\cite{Matsui:2021yte,Martens:2022dtd}.
From these cosmological consequences, the \ac{hl} gravity offers more than merely serving as a candidate for quantum gravity theories.

In this paper, we explore the quantum signature of \ac{pgws} in \ac{hl} gravity. As previously mentioned, \ac{hl} gravity can provide scale-invariant primordial perturbations without inflation, leading one to question how the predictions of \ac{hl} gravity differ from those of inflationary models. Specifically, the generated \ac{pgws} are in a squeezed coherent quantum state, offering a chance to detect their non-classicality by using Hanbury Brown - Twiss (HBT) interferometry. 
The HBT interferometry was originally introduced in the field of radio astronomy~\cite{HanburyBrown:1956bqd,Brown:1956zza}, where it was demonstrated that measuring intensity-intensity correlations could accurately determine the diameters of stars. In quantum optics, these correlations have been used to explore the non-classical properties of photons. This concept was first applied to cosmology in \cite{Giovannini:2010xg}. In this work, we discuss the non-classicality of the \ac{pgws} in \ac{hl} gravity, particularly utilizing the recent approach suggested by Refs~\cite{Kanno:2018cuk,Kanno:2019gqw}.

In particular, we show how the quantum statistics of \ac{pgws} predicted by \ac{hl} gravity differ from those in conventional inflationary models.
Our analysis shows that \ac{pgws} with frequencies higher than $10$ kHz (in a radiation-dominated Universe) or $10^{-3}$ Hz (in a matter-dominated Universe) potentially offer the possibility of detecting their non-classical nature for future observations using HBT interferometry. In contrast, inflationary models require frequencies higher than $10$ kHz~\cite{Kanno:2018cuk,Kanno:2019gqw}. This suggests that while the \ac{pgws} predictions of \ac{hl} gravity in a radiation-dominated Universe and inflationary models are the same in terms of quantum properties, the \ac{hl} predictions in a matter-dominated Universe differ from inflationary predictions. Consequently, this difference is expected to be observable in future experiments.

The rest of the present paper is organized as follows. In Section~\ref{sec:HL-graivty}, we provide a brief review of the construction of the \ac{hl} gravity in $3 + 1$ dimensions. In Section~\ref{sec:quantum-states}, we derive the mode functions of \ac{pgws} in \ac{hl} gravity and introduce squeezed states and squeezed coherent states. We show that \ac{pgws} with squeezed coherent states could arise from the usual interaction between \ac{pgws} and the scalar field. In Section~\ref{sec:graviton-statistics}, we calculate the graviton statistics of the squeezed coherent state and introduce the non-classicality condition of \ac{pgws} detectable using the HBT interferometry. In Section~\ref{sec:possible-detection-nonclassicalPGWs}, we provide calculations of \ac{pgws} generated in radiation- or matter-dominated Universe and predict the frequency range of non-classical \ac{pgws} that could be detected with HBT interferometry. In Section~\ref{sec:conclution-discussion}, we conclude our work.

\section{Ho\v{r}ava-Lifshitz gravity}
\label{sec:HL-graivty}

In this section, we briefly introduce the basic framework of the projectable \ac{hl} gravity. The basic variables are the lapse function $N$, the shift vector $N^i$, and the spatial metric $g_{ij}$ with the positive definite signature $(+,+,+)$. In the \ac{ir}, one can construct the $3 + 1$ dimensional metric out of the basic variables as in the \ac{adm} formalism~\cite{Arnowitt:1962hi}, 
\begin{align}
\mathrm{d} s^2 = -N^2 \mathrm{d} t^2 + g_{i j} (\mathrm{d} x^i + N^i \mathrm{d} t) (\mathrm{d} x^j + N^j \mathrm{d} t)\,,
\end{align}
The shift vector $N^i$ and the spatial metric $g_{ij}$ in general depend on all four coordinates. On the other hand, we consider the projectable \ac{hl} gravity and thus assume that the lapse function $N$ is a function of time only.

The $3 + 1$ dimensional action $S$ describing projectable \ac{hl} gravity is written, in the notation of \cite{Mukohyama:2010xz}, as
\begin{align}\label{HL-action}
S_\textrm{HL}&= \frac{{\cal M}^{2}}{2}\int \mathrm{d} t \mathrm{d}^3 x N \sqrt{g}\,  
\biggl( K^{ij}K_{ij}-\lambda K^2+c^2_gR -2\Lambda+{\mathcal O}_{z>1} \biggr)\,,    
\end{align}
where ${\cal M}$ is the overall mass scale. The extrinsic curvature tensor $K_{ij}$ is defined by $K_{ij}= (\partial_t g_{ij}- g_{j k}\nabla_iN^k-g_{i k}\nabla_jN^k)/(2N)$, with $\nabla_i$ being the spatial covariant derivative compatible with $g_{ij}$, $K^{ij}=g^{ik}g^{jl}K_{kl}$, $K=g^{ij}K_{ij}$ and $R$ is the Ricci scalar of $g_{ij}$, where $g^{ij}$ is the inverse of $g_{ij}$. The constants $\Lambda$ and $c_g$ are the cosmological constant and the propagation speed of tensor gravitational waves, and we set $c_g=1$.
The higher dimensional operators ${\mathcal O}_{z>1}$ is given by
\begin{align}
\frac{{\mathcal O}_{z>1}}{2} & = 
c_1\nabla_iR_{jk}\nabla^iR^{jk}+c_2\nabla_iR\nabla^iR+c_3R_i^jR_j^kR_k^i\nonumber\\
 &  +c_4RR_i^jR_j^i + c_5R^3 
  + c_6R_i^jR_j^i+c_7R^2 \,.
\end{align}
All coupling constants in the action, $\Lambda$ and $c_g$ mentioned above as well as $\lambda$ and $c_{n}$ ($n=1,\cdots,7$) are subject to running under the \ac{rg} flow.

\section{Quantum states of Primordial Gravitational Waves}
\label{sec:quantum-states}

Hereafter,  we will consider a flat \ac{flrw} background and the 
metric is given by
\begin{equation}
\mathrm{d} s^2 = 
-\mathrm{d}t^2+a^2(t)(\gamma_{i j}+h_{i j})\mathrm{d} x^i \mathrm{d} x^j\,,
\end{equation}
where we set $N=1$, $N^i=0$ and $h_{i j}$ is a transverse traceless tensor satisfying $|h_{i j}| \ll \gamma_{i j}$.

The action for the tensor field in a flat \ac{flrw} background is 
given by~\cite{Mukohyama:2009gg}, 
\begin{align}
\begin{split}
S_\textrm{HL}^{(2)}&=
\frac{\mathcal{M}^2}{8} \int \mathrm{d}t \mathrm{d}^3x 
a^3\biggl[\dot{h}^{i j} \dot{h}_{i j}
+ h^{i j}\Delta h_{i j} + 
\left(\frac{1}{\nu^2\mathcal{M}^2}\right)^2 h^{i j}\Delta^3 h_{i j}\biggr]    
\end{split}
\end{align}
where $\Delta=g^{i j}\nabla_i\nabla_j=a(t)^{-2}\gamma^{i j}\nabla_i\nabla_j$ is the Laplacian associated with the spatial metric $g_{i j}$, 
\footnote{
We performed the integral by parts, $-\int \mathrm{d}^3xa^3 \left[g^{kl}\nabla_kh^{ij}\nabla_lh_{ij}\right]
=+\int \mathrm{d}^3xa^3 \left[h^{i j}\Delta h_{i j}\right]$.}
and we have dropped the couplings $c_6$ and $c_7$ for simplicity. We have also introduced a new parameter $\nu$ which is derived from $c_1$ and $c_2$.
Hereafter, we will introduce the conformal time $\eta$ with $\mathrm{d}\eta=\mathrm{d}t/a$,
and the transverse
traceless tensor $h_{ij}(\eta,x^i)$ can be expanded in terms of plane
waves with wavenumber ${\bm k}$ as
\begin{equation}
h_{ij}(\eta, x^i) = \frac{\sqrt{2}}{\mathcal{M}}
\frac{1}{\sqrt{V}}\sum_{\bm k}
\sum_{s} h^s_{\bm{k}}(\eta)\,e^{i {\bm k} \cdot {\bm x}} \ p_{ij}^s(k)  \,,
\label{fourier}
\end{equation}
where $s=\pm$ is the polarization label and 
$p^s_{ij}(k)$ is the polarization tensor normalized as $p^{*s}_{ij} p^{s'}_{ij} =2 \delta^{ss'}$. 
For convenience, we discretized the ${\bm k}$-mode with a width ${\bm k} = (\frac{2\pi n_x}{L_x}, \frac{2\pi n_y}{L_y}, \frac{2\pi n_z}{L_z})$ in a three-dimensional volume $V=L_xL_yL_z$ where ${\bm n}=(n_x,n_y,n_z)$ are integers. The spatial Laplacian leads to $\Delta e^{i {\bm k} \cdot {\bm x}}=a(t)^{-2}k^2e^{i {\bm k} \cdot {\bm x}}$.
By using the above Fourier expansion of $h_{ij}(\eta, x^i)$, we can obtain the following action,
\begin{align}
\begin{split}
S_\textrm{HL}^{(2)}&=\frac{1}{2} \int \mathrm{d}\eta \sum_{\bm k}\sum_{s}
a^2\biggl[h_{\bm k}^{\prime s} h_{\bm{-k}}^{\prime s} -k^2 h_{\bm k}^{s}h_{\bm{-k}}^{s} - 
\left(\frac{1}{\nu^2\mathcal{M}^2}\right)^2 
\frac{k^6}{a^4}h_{\bm k}^{s}h_{\bm{-k}}^{s}\biggr]\,,   
\end{split}
\end{align}
where $k$ is the magnitude of the wave number ${\bm k}$ and $'$ is the derivative with respect to $\eta$. 
By defining $\tilde{h}^s_{\bm{k}}(\eta)=a(\eta) h^s_{\bm{k}}(\eta)$,
the tensor field $\tilde{h}^s_{\bm{k}}(\eta)$ satisfies the equation of motion, 
\begin{equation}
\tilde{h}_{\bm k}^{\prime\prime s}+\left(
\left(\frac{1}{\nu^2\mathcal{M}^2}\right)^2 
\frac{k^6}{a^4}+ k^2-\frac{a''}{a}\right)  \tilde{h}^s_{\bm k}=0\,.
\label{eom}
\end{equation}
In the \ac{uv} limit $\frac{k}{a}\gg 
\nu\mathcal{M}$, Eq.~(\ref{eom}) reduces,
\begin{equation}
\tilde{h}_{\bm k}^{\prime\prime s}+\left(
\frac{k^6}{a^4\nu^4\mathcal{M}^4}-\frac{a''}{a}\right)  \tilde{h}^s_{\bm k}=0\,.
\end{equation}
In this case, we can exactly derive the following mode function, 
\begin{equation}\label{UV-mode-function}
\tilde{h}^s_{\bm k}(\eta)=\frac{\nu\mathcal{M}}{\sqrt{2k^3}}a(\eta)
\exp \left({-i\frac{k^3}{\nu^2\mathcal{M}^2} \int^{\eta}\frac{d\eta'}{a^2(\eta')}}\right)\,,  
\end{equation}
and the corresponding power spectrum is given by 
\begin{equation}
P(k)=k^3|h^s_{\bm k}|^2\propto \nu^2\mathcal{M}^2\,,  
\end{equation}
which means the scale-invariant~\cite{Mukohyama:2009gg}.

\subsection{Squeezed states}

Hereafter, we briefly introduce the Bogoliubov coefficients 
and squeezed operators.
We promote the tensor field $\tilde{h}^s_{\bm{k}}(\eta)$ to the operator, which is expanded as
\begin{align}
\tilde{h}^s_{\bm k}(\eta)=a^s_{\bm k}\,u_k(\eta)+a_{-\bm k}^{s \dagger}\,u_k^{*}(\eta)\,,\ 
\left[a^s_{\bm k} ,a_{\bm p}^{s'\dagger} \right]= \delta^{ss'} \delta_{\bm k,\bm p}\,,
\end{align}
where $*$ denotes complex conjugate.
We can also expand the tensor field $h^s_{\bm{k}}(\eta)$ by $v_{k}(\eta)$ such as
\begin{align}
\tilde{h}^s_{\bm k}(\eta)=b^s_{\bm k}\,v_k(\eta)+b_{-\bm k}^{s \dagger}\,v_k^{*}(\eta)
\,,\ \left[b^s_{\bm k} , b_{\bm p}^{s'\dagger} \right]= \delta^{ss'} \delta_{\bm k,\bm p}\,.
\end{align}

We define the initial vacuum $|0\rangle_{a}$ and late vacuum $|0\rangle_{b}$ 
respectively as 
\begin{equation}
a^s_{\bm k}|0\rangle_{a}=0\,,\qquad
b^s_{\bm k}|0\rangle_{b}=0\,.
\label{vacua}
\end{equation}
The Bogoliubov coefficients $\alpha_k, \beta_k$ are defined as, 
\begin{equation}
v_k^*=\alpha_k u_k^*+\beta_k u_k\,,
\end{equation}
where normalization is kept, if $\left|\alpha_k\right|^2-\left|\beta_k\right|^2=1$. The operators are related via the Bogoliubov transformation:
\begin{align}
b_{\bm k}&=
\alpha_k a_{\bm k}+\beta_k^* a_{-\bm k}^{\dagger}, 
\label{vacuum}\\
b_{\bm k}^{\dagger}&=\alpha_k^* 
a_{\bm k}^{\dagger}+\beta_k a_{-\bm k} .
\end{align}
The standard parametrization of the Bogoliubov coefficients is given
by
\begin{equation}
\alpha_k\equiv\cosh r_k\,, \quad
\beta_k\equiv e^{i\varphi}\sinh r_k\,,
\end{equation}
where the above parametrization is chosen to satisfy the relation $\lvert \alpha_k \rvert^2 - \lvert \beta_k \rvert^2 = 1$.
The squeezing parameter $r_k$ quantifies how much a quantum state is squeezed compared to the vacuum.
The phase factor $e^{i\varphi}$ reflects the complex degree of freedom of $\beta_k$.

By using Eqs.~(\ref{vacua}) and (\ref{vacuum}), the initial vacuum can be given by
\begin{equation}
|0\rangle_{a}
= \prod_{\bm k}\sum_{n=0}^\infty e^{in\varphi}\frac{\tanh^nr_k}{\cosh r_k}\,
|n_{\bm k}\rangle_{b}\otimes|n_{-\bm k}\rangle_{b}\,,
\end{equation}
where $|n_{\bm k}\rangle_{b}=\frac{1}{\sqrt{n!}}\,(b_{\bm k}^{\dagger})^n |0_{\bm k}\rangle_{b}$ and $|0\rangle_{b}=|0_{\bm k}\rangle_{b}\otimes|0_{-\bm k}\rangle_{b}$. It turns out that the initial vacuum is expressed in terms 
of a two-mode squeezed state of the mode $\bm k$ and $-\bm k$ from the viewpoint of the late vacuum. Also, the mean particle density is in proportion to $\left|\beta_k\right|^2$.

\subsection{Squeezed coherent states}

We will now consider the interaction with matter fields to produce coherent states. 
If the \ac{pgws} interact with matter fields perturbatively, coherent states
are generated during the history of the Universe.
In general, the coherent state is given as 
\begin{align}
|\xi_k\rangle_a=e^{-\frac{1}{2}|\xi_k|^2}
\sum_{n=0}^{\infty} \frac{\xi_k^n}{\sqrt{n!}}\,|n_{\bm k}\rangle_a\,.
\end{align}
Now, we introduce the displacement operator
\begin{align}
\hat{D}^{a} (\xi_k ) = \exp\left[\xi_k\,a_{\bm k}^{\dagger} 
- \xi_k^*\,a_{\bm k}\right]\,,
\end{align}
and find the relation,  
\begin{align}\label{coherent}
\hat{D}^{a} (\xi_k )|0\rangle_a 
=|\xi_k\rangle_a\,.
\end{align}
By using the interaction Hamiltonian of the matter field $H_{\rm int}$, we can show that 
the initial state $|0\rangle_{a}$ in the presence of matter fields  becomes a coherent state~\cite{Glauber:1963fi} such as,
\begin{align}
|\xi_k\rangle_a &= \exp\left[\,-i\int d\eta\, H_{\rm int}\,\right] |0\rangle_{a} = \prod_{\bm k} \prod_{s} \hat{D}^{a} (\xi_k )
|0\rangle_{a}\,,
\end{align}
where we will provide an explicit expression below.

We consider that coherent states are generated through the perturbative interaction between \ac{pgws} and a matter field. Although various matter fields could be considered, we introduce a scalar field to simplify the calculations.  We assume a free scalar field $\phi$ in the projectable \ac{hl} gravity, and the action of $\phi$ reads
\begin{align}
S_{\phi}&=\frac{1}{2} \int \mathrm{d} t \mathrm{d}^3 x N \sqrt{g}
\biggl[\dot{\phi}^2 + \phi\Delta \phi + \left(\frac{1}{\nu_\phi^2\mathcal{M}^2}\right)^2 \phi\Delta^3\phi \biggr]\,, 
\end{align}
where $\Delta=g^{i j}\nabla_i\nabla_j$ is the Laplacian associated with the spatial metric $g_{i j}$ and we have dropped the $z=2$ term for simplicity. 
By using the second-order perturbative expansions on the total action of the \ac{hl} gravity, 
we have 
\begin{align}
S^{(2)}&= \int \mathrm{d}t \mathrm{d}^3x 
\Biggl[\frac{\mathcal{M}^2}{8}\biggl(a^3\dot{h}^{i j} \dot{h}_{i j}
+ah^{kl}\left(\partial^i\partial_i\right)h_{kl} +\frac{1}{a^3\nu^4\mathcal{M}^4}h^{kl}\left(\partial^i\partial_i\right)^3h_{kl} \biggr)   \notag
\\
& - \frac{a}{2}\phi\left(h^{ij}\partial_i\partial_j\right)\phi - \left(\frac{3}{2a^{3}\nu_\phi^4\mathcal{M}^4}\right)
\phi\left(h^{ij}\partial_i\partial_j\right)\left(\gamma^{ij}\partial_i\partial_j\right)^2\phi +\cdots
\Biggr]\,.  \end{align}
To construct the displacement operator by using the 
interaction Hamiltonian, we select the terms with two scalars and one tensor. By using the Fourier expansion
of the \ac{pgws} and scalar field, 
we obtain the interaction Hamiltonian for the \ac{pgws}, 
\begin{align}
i\int  &d\eta H_\textrm{int}=
i \int \mathrm{d}\eta \mathrm{d}^3x \biggl[\frac{a^2}{2}\phi\left(h^{ij}\partial_i\partial_j\right)\phi 
+\left(\frac{3}{2a^{2}\nu_\phi^4\mathcal{M}^4}\right)\phi\left(h^{ij}\partial_i\partial_j\right)\left(\gamma^{ij}\partial_i\partial_j\right)^2\phi 
\biggr]  \notag\\
&=i\int\mathrm{d}\eta \,\sum_{\bm k}\sum_{\bm p} \sum_{s}
\biggl\{ \frac{a^2}{2\mathcal{M}}\sqrt{\frac{2}{V}}
\left(p^{ij,s}(\bm{-k})p_ip_j \right)
h^{s}_{-\bm{k}}\phi_{\bm{p}}\phi_{\bm{k-p}} \notag \\
&\quad 
+\frac{3}{2a^{2}\nu_\phi^4\mathcal{M}^5}\sqrt{\frac{2}{V}}
\left(p^{ij,s}(\bm{-k})p_ip_j\right)
\left(\gamma^{ij}p_ip_j\right)^2  h^{s}_{\bm{-k}}\phi_{\bm{p}}
\phi_{\bm{k-p}}\biggr\}\,,    
\end{align}
where we replaced $h^{s}_{\bm{k}}$ with $h^{s}_{-\bm{k}}$ in the last equation for convenience and we expanded $\phi(\eta,x^i)$ in terms of plane waves, 
\begin{equation}
\phi(\eta, x^i) = 
\frac{1}{\sqrt{V}}\sum_{\bm p}
\phi_{\bm{p}}(\eta)\,e^{i {\bm p} \cdot {\bm x}} \,.
\end{equation}
Hence, we can define the interaction Hamiltonian as
\begin{align}
i \int \mathrm{d}\eta\, H_{\rm int}=-\sum_{\bm k}\sum_{s}
\left[\,
\xi_k^s\,a^{s\dagger}_{{\bm k}}-\xi_k^{s*}\,a^s_{\bm k}
\,\right]\,,
\label{interaction1}
\end{align}
where the coherent parameter $\xi_k^s$ is given by
\begin{align}
\xi_k^{s}&=-\frac{i}{\mathcal{M}}\sqrt{\frac{2}{V}}\int\mathrm{d}\eta \sum_{\bm p} \,
\biggl\{\frac{a}{2}\left(p^{ij,s}(\bm{-k})p_ip_j \right)
u_k^{*}(\eta)\phi_{\bm{p}}\phi_{\bm{k-p}}\notag \\
&\quad +
\frac{3}{2a^{3}\nu_\phi^4\mathcal{M}^4}\left(p^{ij,s}(\bm{-k})p_ip_j\right)\left(\gamma^{ij}p_ip_j\right)^2 
u_k^{*}(\eta)\phi_{\bm{p}}\phi_{\bm{k-p}}\biggr\}\,,
\label{interaction-coefficients}      
\end{align}
where $\xi_k^{s}$ is a non-dimensional quantity.
This interaction generates a coherent state such as
\begin{align}
|\xi_k\rangle_a &= \exp\left[\,-i\int d\eta\, H_{\rm int}\,\right] |0\rangle_{a}  \nonumber\\
 &= \prod_{\bm k} \prod_{s} \exp\left[\,
\xi_k^s\,a^{s\dagger}_{{\bm k}}-\xi_k^{s*}\,a^s_{\bm k}
\,\right]
|0\rangle_{a}\,.
\end{align}
Thus, the initial state $|0\rangle_{a}$ in the presence of scalar ﬁelds becomes 
a coherent state~\cite{Glauber:1963fi}.

\section{Graviton statistics and HBT interferometry}
\label{sec:graviton-statistics}

\subsection{Fano factor}

In squeezed vacuum states, the expected number of gravitons is given by
\begin{equation}
 {}_{a}\langle 0| n_{\bm k}|0\rangle_{a}
={}_{a}\langle 0| n_{-\bm k}|0\rangle_{a}=|\beta_k|^2= \sinh^2  r_k\,,
\end{equation}
and the standard variance is computed as
\begin{align}
 \left( \Delta n \right)^2 &=
{}_{a}\langle 0|  \left( n_{\bm k} + n_{- \bm k}  \right)^2  |0\rangle_{a} 
-{}_{a}\langle 0|  n_{\bm k} +n_{- \bm k}  |0\rangle_{a}^2  \notag \\
&=  4 \sinh^2  r_k + 4 \sinh^4 r_k\,,
\end{align}
where we assumed that $n_{\bm k}$ and $n_{-\bm k}$ are indistinguishable and calculated the standard variance for the sum of them.
Then, we can calculate the Fano factor as
\begin{equation}
F = \frac{(\Delta n)^2}{{}_{\rm I}\langle 0|  n_{\bm k} + n_{- \bm k}|0 \rangle_{\rm I}} =  2 + 2 \sinh^2 r_k >1 \,, 
\end{equation}
which shows that the graviton distribution in the squeezed vacuum is super-Poissonian.

Next, let us consider the coherent state.
The expectation number of gravitons can be calculated as
\begin{align}
&{}_{a}\langle \xi_k|n_{\bm k}|\xi_k\rangle_{a}={}_{a}\langle \xi_k|n_{-\bm k}|\xi_k\rangle_{a} 
={}_{b}\langle \xi_k|\hat{S}^\dagger (\zeta)n_{\bm k}\hat{S} (\zeta)
|\xi_k\rangle_{b}\notag \\
&=|\xi_k |^2\left[  e^{-2r_k} \cos^2\left(\theta -\frac{\varphi}{2} \right)
+e^{2r_k}\sin^2 \left(\theta -\frac{\varphi}{2}\right)
\right] +\sinh^2r_k \,,
\end{align}
where $\xi_k = |\xi_k|\,e^{i \theta}$.
The standard variance is also calculated as 
\begin{align}
&(\Delta n)^2={}_{a}\langle \xi_k| \left( n_{\bm k} +n_{- \bm k}\right)^2 
|\xi_k\rangle_{a} 
-{}_{a}\langle \xi_k|n_{\bm k}+n_{-\bm k}|\xi_k\rangle_{a}^2
\notag  \\
&=2|\xi_k |^2\left[e^{-4r_k}\cos^2\left(\theta -\frac{\varphi}{2} \right)
+e^{4r_k}\sin^2\left(\theta -\frac{\varphi}{2}\right)
\right] +4\sinh^2r_k+4\sinh^4r_k\,,
\end{align}
Finally, the Fano factor is found to be
\begin{align}
F &=\frac{|\xi_k |^2e^{-4r_k}+ 2\sinh^2r_k+2\sinh^4r_k}{|\xi_k |^2e^{-2r_k}+\sinh^2r_k}   \,, 
\end{align}
where we consider the case $\theta-\varphi/2=0$.

If the Fano factor $F$ satisfies $F<1$, \ie.
\begin{equation}
|\xi_k |^2 \left(e^{-2r_k} - e^{-4r_k}\right) >\sinh^2 r_k+2\sinh^4 r_k\,,
\label{non-classicality-condition}
\end{equation}
the graviton distribution in the squeezed coherent state can be sub-Poissonian. For $r_k\gg 1$, the condition \eqref{non-classicality-condition} is reduced to 
\begin{equation}
\sinh^6 r_k < \frac{1}{8}|\xi_k|^2\,.
\end{equation}

\subsection{Hanbury Brown - Twiss Interferometry and Its Role in Detecting Primordial Gravitational Waves}

The Hanbury Brown-Twiss (HBT) interferometry, originally developed in radio astronomy to measure stellar angular diameters~\cite{HanburyBrown:1956bqd,Brown:1956zza}, has become a standard technique in quantum optics for probing quantum coherence through intensity-intensity correlations. Its application to \ac{pgws} on its unique ability to detect non-classical signatures in the second-order coherence function, offers a novel pathway to test quantum gravity scenarios in the early Universe~\cite{Kanno:2018cuk,Kanno:2019gqw}.

The HBT interferometry is used to measure intensity-intensity correlations, specifically, the second-order coherence function 
$g^{(2)}(\tau)$, defined as:
\begin{equation}
g^{(2)} (\tau)= \frac{\langle a^\dagger (t) a^\dagger (t+\tau) a (t+\tau)a (t)\rangle}{\langle a^\dagger (t)a (t)\rangle \langle a^\dagger (t+\tau)a (t+\tau)\rangle},
\end{equation}
where $\tau$ represents the time delay between the signals received by the two detectors. Crucially, the HBT does not measure first-order interference but rather intensity fluctuations arising from the quantum statistical properties of the field. For any classical fields, we obtain  $g^{(2)}(0) > 1$ where $g^{(2)}(0)$ is the zero-delay coherence function, while quantum fields (e.g., Squeezed coherent states) can violate this, yielding $g^{(2)}(0) < 1$, a signature of non-classicality.

The zero-delay coherence function $g^{(2)}(0)$ directly connects to the Fano factor $F = (\Delta n)^2 / \langle n \rangle $:
\begin{equation}
g^{(2)} (0) = 1 + \frac{\left( \Delta n \right)^2 - \langle n \rangle}{\langle n \rangle^2}
= 1 + \frac{F - 1}{\langle n \rangle},
\end{equation}
where $\langle n \rangle$ and $(\Delta n)^2$ are the mean graviton number and its variance, respectively. 
Therefore, $g^{(2)} (0)$ will be less than one when the Fano factor is below one, signaling non-classicality. If the early Universe went through a phase where the Fano factor was less than one, it could allow us to observe a non-classical signature in \ac{pgws}  using the HBT interferometry.

\section{Possible detection of non-classical PGWs 
in Ho\v{r}ava-Lifshitz cosmology}
\label{sec:possible-detection-nonclassicalPGWs}

\subsection{Radiation-dominated Universe}

Hereafter, we will focus on the \ac{hl} gravity.
For simplicity, we assumed the mode function changes 
from the anisotropic (\ac{uv}) regime $\frac{k}{a} > \nu\mathcal{M}$ to 
ordinary (\ac{ir}) regime $\frac{k}{a}< \nu\mathcal{M}$. 
First, we assume the radiation-dominated Universe in which 
the scale factor is written as
\begin{equation}
a(\eta)=C_{r}\eta,  \qquad 0<\eta<\eta_{\rm eq},
\end{equation}
where $\eta_{\rm eq}$ is the conformal time of the matter-radiation equality.
Since we are considering the radiation-dominated Universe, we have to be careful whether it is a thermal vacuum or not. However, the gravitational interaction is extremely weak and there is no need to consider a graviton thermal bath as long as the Planck temperature is not reached. Thus, in this paper, we do not consider the thermal vacuum~\cite{Koh:2004ez}.
Then, Eq.~\eqref{eom} gives the positive frequency mode in the \ac{uv} and \ac{ir} regimes respectively as
\begin{align}
\left\{
\begin{array}{l}
\vspace{0.2cm}
u_{k}(\eta) =\frac{\nu\mathcal{M}}{\sqrt{2k^3}}
C_{r}\eta\exp \left({i\frac{k^3}{\nu^2\mathcal{M}^2C_{r}^2\eta}}\right)\,,\\
\vspace{0.2cm}
v_{k}(\eta)\equiv\frac{1}{\sqrt{2k}}\,e^{-ik \eta } \,,
\end{array}
\right. 
\label{modefunction-radiation-dominated}
\end{align}
where we have used Eq.~\eqref{UV-mode-function} as the mode function in the \ac{uv} regime.

We introduce the conformal time parameter $\eta_1$ at which the Universe transitions from the \ac{uv} regime with the anisotropic scaling, where the dispersion relation is dominated by the higher-order $k^6$ term, to the \ac{ir} regime, where the usual quadratic $k^2$ dispersion relation dominates. Specifically, we match the \ac{uv} solution given by $u_{k}$ of Eq.~\eqref{modefunction-radiation-dominated}, which captures the mode function valid at early times (respecting the anisotropic scaling), to the \ac{ir} solution given by $v_{k}$ of Eq.~\eqref{modefunction-radiation-dominated}. This matching occurs precisely at $\eta = \eta_1$, satisfying the continuity conditions for both mode functions and their first derivatives.
The corresponding Bogoliubov coefficients read, 
\begin{align}
\alpha_k=\frac{e^{2ik\eta _1} \left(2k\eta _1+i\right)}{2k\eta _1},\quad
\beta_k= -\frac{i}{2k\eta _1}\,,
\end{align}
and the squeezing parameter is written by
\begin{equation}\label{squeezing-parameter-hl-radiation}
\sinh r_k
=\biggl|\frac{1}{2 \eta _1 k}\biggr|\,,
\end{equation}
where the matching condition at $\eta=\eta_1$ leads to the relation $\nu\mathcal{M}=\frac{k}{C_{r}\eta_{1}}$.

We translate the comoving wave number $k$ into the physical frequency at present~\cite{Maggiore:1999vm}. The physical frequency is written by 
\begin{equation}
2\pi f= \frac{k}{a(t_0)}\,,
\end{equation}
where $t_0$ is the present value of cosmic time and we shall set $a(t_0)=1$. We estimate the quantity $k\eta_1$ and obtain the following relation, 
\begin{align}
k\eta_1=2\pi fa(t_0)\eta_1
=\frac{\left(2\pi fa(t_0)\right)^2}{\nu\mathcal{M}C_{r}}\,,
\end{align}
where we used $a(\eta)=C_{r}\eta $ and $\frac{k}{a(\eta_{1})}=\nu\mathcal{M}$.
By using the following relation
\begin{equation}
a(t_0)=\sqrt{2C_{r}}\left(\frac{t_0}{t_{\rm eq}}\right)^{2/3}t_{\rm eq}^{1/2}\,,
\end{equation}
we can obtain the following expression
\begin{align}
k\eta_1=\frac{2(2\pi f)^2}{\nu\mathcal{M}}
\left(\frac{t_0^{2/3}}{t_{\rm eq}^{2/3}}t_{\rm eq}^{1/2}\right)^{2}
\equiv \left(\frac{f}{f_1}\right)^2\,,
 \label{eqn:keta1}
\end{align}
where 
\begin{align}
f_1 =
\frac{1}{2\pi}\frac{t_{\rm eq}^{1/6}}{t_0^{2/3}} \sqrt{\frac{\nu\mathcal{M}}{2}}
\simeq 10^9 \sqrt{\frac{\nu\mathcal{M}}{10^{-4}M_{\rm pl}}}\quad[{\rm Hz}]\,,
\end{align}
where we introduced the Planck mass $M_{\rm pl}^2=\frac{1}{8\pi G}$.

When the system was in the vacuum state before the de Sitter-radiation transition, after the transition the mean particle density $N_f$ is given by
\begin{equation}
N_f= \left|\beta_k\right|^2 =\frac{1}{4}\left(\frac{f_1}{f}\right)^4\,.
\end{equation}
By using the above expression of $N_f$
we have the amplitude of the \ac{pgws},
\begin{align}
h_0^2 \Omega_{\mathrm{gw}}(f) \simeq 3.6\left(\frac{N_f}{10^{37}}\right)\left(\frac{f}{1 \mathrm{kHz}}\right)^4 \simeq 10^{-13}\left(\frac{\nu\mathcal{M}}{10^{-4}M_{\rm pl}}\right)^2\,,
\end{align}
which is the same as the inflation prediction with $H_{\rm inf}$ replaced by $\nu\mathcal{M}$~\cite{Maggiore:1999vm}.

Plugging \eqref{eqn:keta1} back into Eq.~\eqref{squeezing-parameter-hl-radiation}, we have
\begin{align}
\sinh r_k=\frac{1}{2}\left(\frac{f_1}{f}\right)^2\,,
\end{align}
which is the same form as the inflation models~\cite{Maggiore:1999vm}.
Hereafter, we drop the spin degree of freedom of $|\xi_k^{s}|$ for simplicity.
Combining the above relation with the condition~\eqref{non-classicality-condition}, 
we obtain the condition to observe non-classical \ac{pgws} where we assume $r_k\gg 1$, 
\begin{align}\label{nonclassical-condition-radiation}
f  > \left(\frac{1}{8}\right)^{\frac{1}{12}}\,|\xi_k |^{-\frac{1}{6}}f_1 
= \left(\frac{1}{8}\right)^{\frac{1}{12}}10^9\,|\xi_k |^{-\frac{1}{6}}
\sqrt{\frac{\nu\mathcal{M}}{10^{-4}M_{\rm pl}}}\quad[{\rm Hz}] \,.
\end{align}
The above condition~\eqref{nonclassical-condition-radiation} 
describes the condition under which the graviton distribution in the squeezed coherent state becomes sub-Poissonian (Fano factor $F < 1$), indicating non-classicality. We note that this condition is consistent with the standard prediction of the inflation~\cite{Kanno:2018cuk,Kanno:2019gqw} when we replace the 
mass scale $\nu\mathcal{M}$ into the inflationary Hubble expansion rate $H_{\rm inf}$.

Now, let us estimate the coherent parameter 
$|\xi_k|$ for the radiation-dominated Universe. 
By using Eq~\eqref{interaction-coefficients}, we obtain 
\begin{align}
\xi_k&=-\frac{i}{\mathcal{M}}\sqrt{\frac{2}{V}}\int_{\eta_0}^{\eta_1}\mathrm{d}\eta \sum_{\bm p}\,
\biggl\{\frac{a(\eta)}{2}p^2
u_k^{*}(\eta)\phi_{\bm{p}}\phi_{\bm{k-p}}\notag \\
&\quad + \frac{3}{2a(\eta)^{3}\nu_\phi^4\mathcal{M}^4}p^6
u_k^{*}(\eta)\phi_{\bm{p}}\phi_{\bm{k-p}}\biggr\}\,,    
\end{align}
where we can approximately use Eq.~\eqref{UV-mode-function} 
as the mode functions of the \ac{pgws} and scalar field.
By using the following expressions, 
\begin{align}
u_k^{*}(\eta)&
=\frac{\nu\mathcal{M} }{\sqrt{2k^3}}a(\eta)
e^{+i\frac{k^3}{\nu^2\mathcal{M}^2C_{r}^2}\left[\frac{1}{\eta_0}-\frac{1}{\eta_1}\right]}\,, \\
\phi_{p}&
=\frac{\nu_\phi\mathcal{M} }{\sqrt{2p^3}}
e^{-i\frac{p^3}{\nu_\phi^2\mathcal{M}^2C_{r}^2}\left[\frac{1}{\eta_0}-\frac{1}{\eta_1}\right]}\,, \\
\phi_{k-p}&=\frac{\nu_\phi\mathcal{M} }{\sqrt{2(k-p)^3}}
e^{-i\frac{(k-p)^3}{\nu_\phi^2\mathcal{M}^2C_{r}^2}\left[\frac{1}{\eta_0}-\frac{1}{\eta_1}\right]}\,,
 \end{align}
we obtain
\begin{align}
\xi_k&=-\frac{i}{\mathcal{M}}\sqrt{\frac{2}{V}}\int_{\eta_0}^{\eta_1}\mathrm{d}\eta \frac{\nu\nu_\phi^2\mathcal{M}^3 }{\sqrt{2k^3}}\sum_{\bm p}\,
\biggl\{\frac{a(\eta)^2}{4}\frac{p^{\frac{1}{2}}}{\sqrt{(k-p)^3}} \notag \\
&\quad + \frac{3}{4a(\eta)^{2}{\nu_\phi^4}\mathcal{M}^4}
\frac{p^{\frac{9}{2}}}{\sqrt{(k-p)^3}}\biggr\}e^{+\frac{i}{\mathcal{M}^2C_{r}^2}\left[\frac{1}{\eta_0}-\frac{1}{\eta_1}\right]\left(\frac{k^3}{\nu^2}-\frac{p^3}{\nu_\phi^2}-\frac{(k-p)^3}{\nu_\phi^2}\right)}\,,  
\end{align}
where it may be important to determine whether we can drop the exponential factor. The exponential factor represents rapidly oscillating phases arising from interactions during the transition period. Physically, if this factor is not negligible, the generation of squeezed coherent states may be strongly suppressed, depending on the mode. Therefore, the condition of eliminating this factor guarantees the efficient generation of squeezed coherent states.
Conservatively, we should take $\eta_0\sim \eta_1$ and $\nu \sim \nu_\phi$.
By using the matching condition $\nu\mathcal{M}=\frac{k}{C_{r}\eta_{1}}$, we obtain the condition under which the exponential factor can be dropped as
\begin{equation}
\frac{k^3}{\nu^2\mathcal{M}^2C_{r}^2}\frac{1}{\eta_1}\sim k\eta_{1} = \left(\frac{f}{f_1}\right)^2
=\left(\frac{f}{10^9 \sqrt{\frac{\nu\mathcal{M}}{10^{-4}M_{\rm pl}}}\ [{\rm Hz}]}\right)^2 \ll 1\,.
\end{equation}
For $f < 10^{5}$ and reasonable values of $\nu\mathcal{M}$, this condition holds and we can safely neglect the exponential factor.

For a large three-dimensional volume $V$ the mode $\bm{p}$  
for the scalar field is treated as continuous, $
\frac{1}{V}\sum_{\bm p} \to \int \frac{\mathrm{d}^3{p}}{(2\pi)^3}$.
Thus, we obtain approximately
\begin{align}
\xi_k&=-\frac{i\sqrt{2V}}{\mathcal{M}}\int_{\eta_0}^{\eta_1}\mathrm{d}\eta \frac{\nu\nu_\phi^2\mathcal{M}^3 }{\sqrt{2k^3}}\int_{p\,\sim\,k}  \frac{\mathrm{d}^3{p}}{(2\pi)^3}
\biggl\{\frac{a(\eta)^2}{4}\frac{p^{\frac{1}{2}}}{\sqrt{(k-p)^3}} 
+ \frac{3}{4a(\eta)^{2}{\nu_\phi^4}\mathcal{M}^4}
\frac{p^{\frac{9}{2}}}{\sqrt{(k-p)^3}}\biggr\} \notag \\
&\simeq -\frac{i\nu\nu_\phi^2\mathcal{M}^2}{(2\pi)^3}
\sqrt{\frac{V}{k^3}}\int_{\eta_0}^{\eta_1}\mathrm{d}\eta
\biggl\{\frac{a(\eta)^2}{4}k^2 
+ \frac{3k^6}{4a(\eta)^{2}{\nu_\phi^4}\mathcal{M}^4}\biggr\}\,,
\end{align}
where we approximate $\int_{p\,\sim\,k} \mathrm{d}^3{p}\sim k^3$.
Now, we obtain the following estimate, 
\begin{align}\label{coherent-parameter-radiation}
|\xi_k| 
&\simeq \frac{k^{\frac{3}{2}}\sqrt{V}}{96\pi^3}\frac{\nu_\phi^2}{\nu}(k\eta_1)+ 
\frac{3k^{\frac{3}{2}}\sqrt{V}}{32\pi^3}\frac{\nu^3}{\nu_\phi^2}(k\eta_1)\left(\frac{\eta_1}{\eta_0}\right)\,,   
\end{align}
where we take $\eta_1 > \eta_0$.

First, we will assume the case where the tree interaction is dominant and consider only the first term in Eq.~\eqref{coherent-parameter-radiation}. Utilizing the non-classicality condition~\eqref{nonclassical-condition-radiation},
we obtain 
\begin{align}
f  &> \left(\frac{144\pi^3}{V}\right)^{\frac{1}{19}}\left(\frac{\nu}{\nu_\phi^2}\right)^{\frac{2}{19}}
f_1^{\frac{16}{19}} 
 = 8.5\times 10^{4}\left(\frac{\nu}{\nu_\phi^2}\right)^{\frac{2}{19}}\left(\frac{\nu\mathcal{M}}{10^{-4}M_{\rm pl}}\right)^{\frac{8}{19}}
\quad[{\rm Hz}] \,,
\end{align}
where we take $V= H_0^{-3}$ with the current Hubble constant $H_0$.
Next, we will examine the case where the second term, characterized by the higher-order interaction of the scalar field and \ac{pgws}, is dominant. Utilizing the non-classicality condition~\eqref{nonclassical-condition-radiation}, we obtain
\begin{align}
f & > \left(\frac{16\pi^3}{9V}\right)^{\frac{1}{19}}\left(\frac{\nu_\phi^2}{\nu^3}\right)^{\frac{2}{19}}\left(\frac{\eta_0}{\eta_1}\right)^{\frac{2}{19}}
f_1^{\frac{16}{19}} \notag \\
&= 7.2\times 10^{4}\left(\frac{\nu_\phi^2}{\nu^3}\right)^{\frac{2}{19}}\left(\frac{\eta_0}{\eta_1}\right)^{\frac{2}{19}}\left(\frac{\nu\mathcal{M}}{10^{-4}M_{\rm pl}}\right)^{\frac{8}{19}}
\quad[{\rm Hz}] \,.
\end{align}
Conservatively, we set $\nu \sim \nu_\phi \sim 10^{-4}$ and $\mathcal{M} \sim M_{\rm pl}$. 
\footnote{ 
The parameters $\nu\mathcal{M}, \nu_\phi\mathcal{M}$ 
correspond to the amplitude of fluctuations~\cite{Mukohyama:2009gg}, which similar to inflationary models, is constrained observationally, i.e., $H_{\rm inf}\lesssim 10^{-4}M_{\rm pl}$. Hence, choosing $\nu \sim \nu_\phi \sim 10^{-4}$ is conservative because it aligns with typical amplitude constraints derived from inflationary scenarios. }
Additionally, while $(\eta_0/\eta_1) < 1$, a conservative estimate would be $(\eta_0/\eta_1)^{\frac{2}{19}}\sim \mathcal{O}(1)$. However, when $\eta_0$ is significantly smaller than $\eta_1$, i.e. if the \ac{pgws} and scalar field can interact sufficiently, the frequency range where quantum effects can be observed becomes lower. Interestingly, if the effect of $(\eta_0/\eta_1)^{\frac{2}{19}}$ is ignored, the quantum nature constraints on tree interaction and higher-order interaction of the scalar field and \ac{pgws} are of the same order.
Regarding the frequency and quantum nature of \ac{pgws} like the squeezing parameter, there are no differences between \ac{hl} gravity in the radiation-dominated Universe and inflationary models~\cite{Watanabe:2009ct,Soda:2012zm,Barnaby:2010vf}. Although the coherent parameter $|\xi_k|$ derived from the interaction Hamiltonian of scalar fields and \ac{pgws} in \ac{hl} gravity theory differs from that in inflation theory, the observational quantum signature of \ac{pgws} based on \ac{hl} gravity are the same as the inflationary predictions. In the radiation-dominated Universe, detecting the non-classicality of \ac{pgws} using the HBT interferometry requires a frequency higher than $10$ kHz, which is the same order as inflationary predictions~\cite{Kanno:2018cuk,Kanno:2019gqw}. 
However, the predictions of \ac{hl} gravity in the radiation-dominated Universe are unique. If the cosmic expansion differs from that expected in the radiation-dominated Universe, this constraint may also change. Indeed, in the next section, we shall consider the matter-dominated Universe, and show the condition for the detectability of non-classicality is notably weaker.

\subsection{Matter-dominated Universe}

Next, we consider the \ac{pgws} in the matter-dominated Universe where 
the scale factor is written as
\begin{equation}
a(\eta)=C_{m}\eta^2  \,.
\end{equation}
We can get the non-classicality condition for the matter-dominated Universe scenario as the same as the radiation-dominated Universe.

Then, Eq.~(\ref{eom}) gives the positive frequency mode 
at \ac{uv} or \ac{ir} regime as
\begin{align}
\left\{
\begin{array}{l}
\vspace{0.2cm}
u_{k}(\eta)=\frac{\nu\mathcal{M}}{\sqrt{2k^3}}
C_{m}\eta^2\exp \left({i\frac{k^3}{3C_{m}^2\nu^2\mathcal{M}^2\eta^3}}\right)\,,\\
\vspace{0.2cm}
v_{k}(\eta)= \frac{1}{\sqrt{2k}}\left(1-\frac{i}{k \eta}\right)e^{-ik\eta}\,.
\end{array}
\right. 
\label{modefunction-matter-dominated}
\end{align}
The corresponding Bogoliubov coefficients read, 
\begin{align}
\alpha_k=\frac{e^{\frac{4 ik\eta_2}{3}} (-3+2k\eta_2 (k\eta_2+2i))}{2 k^2\eta_2^2},\quad 
\beta_k= \frac{e^{\frac{2 ik\eta_2}{3}} (3-2 ik\eta_2)}{2k^2\eta_2^2}\,,
\end{align}
and the squeezing parameter reads, 
\begin{equation}\label{squeezing-parameter-hl-matter}
\sinh r_k=\biggl|\frac{3-2ik\eta_1}{2k^2\eta_1^2}\biggr|\,,
\end{equation}
where we used the matching condition at $\eta=\eta_2$ with $\nu\mathcal{M}=\frac{k}{C_{m}\eta^2_{2}}$.

We estimate the quantity $k\eta_2$ and obtain the following relation, 
\begin{align}
k\eta_2=2\pi fa(t_0)\eta_2
=\frac{\left(2\pi fa(t_0)\right)^{\frac{3}{2}}}{\nu^{\frac{1}{2}}\mathcal{M}^{\frac{1}{2}}C_{m}^{\frac{1}{2}}}\,,
\end{align}
where we used $a(\eta)=C_{m}\eta^2 $ and $\frac{k}{a(\eta_{2})}=\nu\mathcal{M}$.
By using the relation
\begin{equation}
a(t_0)=C_{m}^{1/3}\left(3t_0\right)^{2/3}\,,
\end{equation}
we obtain the following expression
\begin{align}
k\eta_2=\frac{\left(2\pi f\right)^{\frac{3}{2}}(3t_0)}{
\nu^{\frac{1}{2}}\mathcal{M}^{\frac{1}{2}}}\equiv 
\left(\frac{f}{f_2}\right)^{\frac{3}{2}}\,, 
\end{align}
where $f_2$ is given by
\begin{equation}
f_2=\frac{1}{2\pi}\left(\frac{\nu\mathcal{M}}{9t_0^2}\right)^{\frac{1}{3}}=
\left(\frac{\nu\mathcal{M}}{10^{-4}M_{\rm pl}}\right)^{\frac{1}{3}}\ [{\rm Hz}] \,.
\end{equation}
Plugging this back into Eq.~\eqref{squeezing-parameter-hl-matter}, we get
\begin{align}
\sinh r_k\simeq \frac{3}{2k^2\eta_2^2}= \frac{3}{2}\left(\frac{f_2}{f}\right)^3\,,
\end{align}
where we assume $k\eta_2\ll 1$.
Combining the above relation with the condition~\eqref{non-classicality-condition}, 
we obtain the condition to observe non-classicality \ac{pgws}~\cite{Kanno:2018cuk,Kanno:2019gqw} where we assume $r_k\gg 1$, 
\begin{align}\label{nonclassical-condition-matter}
f  > \left(\frac{9}{2}\right)^{\frac{1}{6}}\,|\xi_k |^{-\frac{1}{9}}f_2 
=\left(\frac{9}{2}\right)^{\frac{1}{6}}\,|\xi_k |^{-\frac{1}{9}}
\left(\frac{\nu\mathcal{M}}{10^{-4}M_{\rm pl}}\right)^{\frac{1}{3}}\ [{\rm Hz}]\,.
\end{align}

Now, let us estimate the coherent parameter $|\xi_k|$ 
for the matter-dominated Universe. As in the calculation for the case of the radiation-dominated Universe, by using Eq~\eqref{interaction-coefficients}, we obtain 
\begin{align}
\xi_k &=-\frac{i}{\mathcal{M}}\sqrt{\frac{2}{V}}\int_{\eta_0}^{\eta_2}\mathrm{d}\eta \frac{\nu\nu_\phi^2\mathcal{M}^3 }{\sqrt{2k^3}}\sum_{\bm p}\,
\biggl\{\frac{a(\eta)^2}{4}\frac{p^{\frac{1}{2}}}{\sqrt{(k-p)^3}} \notag \\
&\quad + \frac{3}{4a(\eta)^{2}{\nu_\phi^4}\mathcal{M}^4}
\frac{p^{\frac{9}{2}}}{\sqrt{(k-p)^3}}\biggr\}e^{+\frac{i}{3\mathcal{M}^2C_{m}^2}\left[\frac{1}{\eta_0^3}-\frac{1}{\eta_2^3}\right]\left(\frac{k^3}{\nu^2}-\frac{p^3}{\nu_\phi^2}-\frac{(k-p)^3}{\nu_\phi^2}\right)}\notag \\
&\simeq -\frac{i\nu\nu_\phi^2\mathcal{M}^2}{(2\pi)^3}
\sqrt{\frac{V}{k^3}}\int_{\eta_0}^{\eta_2}\mathrm{d}\eta
\biggl\{\frac{a(\eta)^2}{4}k^2
+ \frac{3k^6}{4a(\eta)^{2}{\nu_\phi^4}\mathcal{M}^4}\biggr\}\,,
\end{align}
where we have dropped the exponential factor in the last equation as in the previous radiation case.
Now, we obtain the following expression, 
\begin{align}
|\xi_k| 
&\simeq \frac{k^{\frac{3}{2}}\sqrt{V}}{160\pi^3}\frac{\nu_\phi^2}{\nu}(k\eta_2)+ \frac{k^{\frac{3}{2}}\sqrt{V}}{32\pi^3}\frac{\nu^3}{\nu_\phi^2}(k\eta_ 2)\left(\frac{\eta_2}{\eta_0}\right)^3\,,   
\end{align}
where we have used $\frac{k}{C_{m}\eta_2^2}=\nu\mathcal{M}$, 
and taken $\eta_2 > \eta_0$.

First, we will consider the case where the tree interaction is dominant and the first term in the above expression is dominant. Utilizing the non-classicality condition~\eqref{nonclassical-condition-matter},
we obtain 
\begin{align}
f  &> \left(\frac{291600\pi^3}{V}\right)^{\frac{1}{24}}\left(\frac{\nu}{\nu_\phi^2}\right)^{\frac{1}{12}}
f_2^{\frac{7}{8}} 
 = 1.2\times 10^{-2}\left(\frac{\nu}{\nu_\phi^2}\right)^{\frac{1}{12}}\left(\frac{\nu\mathcal{M}}{10^{-4}M_{\rm pl}}\right)^{\frac{7}{24}}
\ [{\rm Hz}] \,.
\end{align}
Next, we will examine the case where the second term, characterized by the higher-order interaction of the scalar field and \ac{pgws}, is dominant. Utilizing the non-classicality condition~\eqref{nonclassical-condition-matter}, we obtain 
\begin{align}
f & > \left(\frac{11664\pi^3}{V}\right)^{\frac{1}{24}}
\left(\frac{\nu_\phi^2}{\nu^3}\right)^{\frac{1}{12}}
\left(\frac{\eta_0}{\eta_2}\right)^{\frac{1}{4}}f_2^{\frac{7}{8}} \notag \\
&= 1.0\times 10^{-2}\left(\frac{\nu_\phi^2}{\nu^3}\right)^{\frac{1}{12}}
\left(\frac{\eta_0}{\eta_2}\right)^{\frac{1}{4}}\left(\frac{\nu\mathcal{M}}{10^{-4}M_{\rm pl}}\right)
^{\frac{7}{24}}
\quad[{\rm Hz}] \,.
\end{align}
Compared to the radiation-dominated Universe, the frequency of \ac{pgws} required to detect non-classicality using the HBT interferometry is significantly reduced to $f\gtrsim 10^{-3}$ Hz with a conservative set-up. Specifically, the lowest frequency required for the detectability of non-classicality decreases by a factor of $10^{-7}$ in comparison with the radiation-dominated Universe. Conservatively, the non-classicality of \ac{pgws} generated during the matter-dominated Universe can be detected using the LISA detector.

\section{Conclusion and discussions}
\label{sec:conclution-discussion}

In this paper, we have investigated the quantum signature of \ac{pgws} generated in \ac{hl} gravity. This theory uniquely allows for the generation of scale-invariant primordial perturbations without inflation, while satisfying the renormalizability, unitarity and asymptotic freedom as a quantum field theory of gravity.
We have focused on examining the non-classical nature of \ac{pgws} in \ac{hl} gravity and the potential for detecting this non-classicality.

The main findings are as follows. First, we have demonstrated that \ac{hl} gravity can generate scale-invariant \ac{pgws} during both the radiation-dominated and matter-dominated Universe, as originally suggested in \cite{Mukohyama:2009gg}. We have analytically constructed the mode functions for \ac{uv} and \ac{ir} modes in these stages of the Universe. Second, we have shown that for the radiation-dominated Universe, the squeezing parameter takes the same form as in the inflationary model, but for the matter-dominated Universe, it does not. 
Third, when scalar fields are present, the squeezed quantum state of \ac{pgws}  transitions into a coherent state. We calculated the coherent parameter that arises from the interaction between \ac{pgws}  and the scalar field in \ac{hl} gravity. Owing to the higher-dimensional operators characteristic of \ac{hl} gravity, both the amplitude of the \ac{pgws}  and the scalar field, as well as the coherent parameter, become substantially enhanced. This enhancement opens the possibility of experimentally probing the non-classicality of \ac{pgws}  through HBT interferometry. Specifically, we have shown that detecting \ac{pgws} with frequencies higher than $10$ kHz (if generated in the radiation-dominated Universe) or $10^{-3}$ Hz (if generated in the matter-dominated Universe) enables us to detect their non-classicality with HBT interferometry, whereas inflationary models require frequencies higher than $10$ kHz~\cite{Kanno:2018cuk,Kanno:2019gqw}.

In this paper, we explored the possibility of testing the quantum nature of the \ac{pgws} generated in the radiation-dominated Universe and matter-dominated Universe. While the predictions for the radiation-dominated Universe were found to be the same as those of inflationary models, the lowest frequency required for the detectability of non-classicality of the \ac{pgws} may be relaxed by more general cosmic expansion, as demonstrated in the matter-dominated Universe. Since the \ac{hl} gravity can generally generate scale-invariant \ac{pgws} in any expansion of the early Universe $a \propto t^n$ with $n>1/3$~\cite{Mukohyama:2009gg}, it is expected that the condition for the detectability of non-classicality may be sufficiently weaker, and we will deal with this in future work. Our results indicate that \ac{hl} gravity not only serves as a candidate for quantum gravity theories but also offers significant cosmological predictions. Particularly, the distinct squeezing and coherent parameters in \ac{hl} gravity suggest different observational outcomes for the quantum signature of \ac{pgws} compared to inflationary models~\cite{Watanabe:2009ct,Soda:2012zm,Barnaby:2010vf}.

\section*{Acknowledgments}
We thank Jiro Soda for useful discussion and comments.
S.~K. was supported by the Japan Society for the Promotion of Science (JSPS) KAKENHI Grant Numbers JP22K03621, JP22H01220, 24K21548 and MEXT KAKENHI Grant-in-Aid for Transformative
Research Areas A “Extreme Universe” No. 24H00967. H.~M. was supported by JSPS KAKENHI Grant No. JP22KJ1782 and No. JP23K13100.
S.~M was supported in part by Japan Society for the Promotion of Science (JSPS) Grants-in-Aid for Scientific Research No.~24K07017 and the World Premier International Research Center Initiative (WPI), MEXT, Japan.

\bibliography{Refs}
\bibliographystyle{JHEP}

\end{document}